\documentclass[12pt,a4paper,final]{article}
\usepackage[dvipsnames]{xcolor}
\usepackage{color}
\usepackage[english]{babel}
\usepackage{amsmath}
\usepackage{amsfonts}
\usepackage{amssymb}
\usepackage{graphicx}
\def\ai{\'{\i}}
\def\lp{\left(}
\def\rp{\right)}

\def\be{\begin{equation}}
\def\ee{\end{equation}}

\begin{document}
\title{\vspace{-1.5cm} \bf Tides across thin-shells: differences between spacetimes with one and two asymptotic regions}
\author{E. Rub\ai n de Celis\footnote{e-mail: erdec@df.uba.ar} and C. Simeone\footnote{e-mail: csimeone@df.uba.ar}\\
{\footnotesize 1. Universidad de Buenos Aires, Facultad de Ciencias Exactas y Naturales,}\\ {\footnotesize Departamento de F\ai sica, Buenos Aires, Argentina }\\
{\footnotesize 2. CONICET,  Universidad de Buenos Aires, Instituto de F\ai sica de Buenos Aires (IFIBA),}\\
{\footnotesize Buenos Aires, Argentina.}}
\date{\small \today}

\maketitle
\vspace{0.6cm} 
\begin{abstract}

Traversability across thin shells is
investigated, with special attention devoted to the difference in tides related with different global properties of the geometries. While  we have recently associated curvature jumps across infinitely thin shells to troublesome tides and consequent very restrictive conditions for a safe travel across a throat satisfying the flare-out condition in spacetimes with two asymptotic regions, now we find that analogous problems   can be significantly reduced or even cancelled across shells joining an inner with an outer submanifold of spacetimes with one asymptotic region. We also show that, within this framework, cylindrical shells present an advantage over spherical shells.

\vspace{1cm} 

\noindent 
PACS number(s): 4.20.-q, 04.20.Gz, 04.40.Jb\\


\end{abstract}

\vspace{1cm}

\section{Introduction}

Our recent study in \cite{nos21,nos22} shows that the central problem with tides across thin shells at the throat of geometries with two asymptotic regions comes from the jump in the components of the extrinsic curvature tensor; the behaviour of the curvature in the nearby smooth regions does not pose, instead, the same kind of difficulty (see for instance \cite{martin} and, for a related analysis, the recent work \cite{genc}). For the radial tides on radially moving objects this curvature jump contribution is fixed even for infinitely close points; hence we would have an, in principle, divergent quotient $\Delta a/\Delta x$  of the relative acceleration and the separation between two points at different sides of the shell. Angular tides on radially moving objects could seem to avoid this kind of problem because the contribution of the curvature jump is proportional to the angular transverse separation; however, the jump turns out to appear in  a quotient by the travelling time across the shell which, formally, is infinitely short.  In the case of axially symmetric problems, the longitudinal transverse tide presents the same peculiarity. Then a safe passage through such configurations would be jeopardized by, in principle, insurmountable tides. However, in our more relaxed approach relying on understanding those formal results as a first approximation to more realistic shells with a little but non vanishing thickness, it becomes a natural idea the search for conditions which could reduce the jump of the components of the extrinsic curvature across the shell. In particular, in \cite{nos22} we have studied the advantages provided by wormhole constructions within theories beyond general relativity. Here we extend our study to the problem of tides across thin shells in spacetimes with only one asymptotic region; we show that important advantages are provided by the additional freedom available when thin matter layers connect inner to outer submanifolds at a surface where a flare-out condition is not satisfied\footnote{Such condition means that the geodesics open up at a given surface; in its usual form it corresponds to the existence of a minimal area surface ({\it areal} definition), but for cylindrically symmetric geometries  an alternative less restrictive definition ({\it circular} one \cite{brle}) has also been proposed involving just the geodesics on a plane normal to the symmetry axis (see below).}. In such more physically appealing cases the different signs in the terms involved allow for parameter choices significantly reducing or even eliminating the jump of the extrinsic curvature; hence the corresponding troublesome contributions to tides across the shells could be avoided. We also show the important differences between spherically symmetric problems and cylindrically symmetric ones, as in the latter not only the problematic radial tide can be managed, but also the transverse tides could be reduced or eliminated without resorting to matter with non standard properties.

\section{Tides across thin-shells}

We consider only highly symmetric static geometries connected at a spherical or cylindrical surface $\Sigma$ where a --formally-- infinitely thin shell is placed; introducing the sign $\pm$ to denote each side of the shell, the corresponding line elements are
\be\label{metric1}
ds_{\pm}^2 = g^{\pm}_{00}\, dt^2 + g^{\pm}_{rr}\, dr_{\pm}^2 + g^{\pm}_{\zeta\zeta}\, d\zeta^2 + g^{\pm}_{\varphi\varphi}\, d\varphi^2 \,.
\ee
The radial coordinates satisfy $r_\pm\geq b_\pm$ for spacetimes with  two asymptotic regions (wormholes), and $0\leq r_-\leq b_-,\ b_+\leq r_+$ for only one asymptotic region. The coordinate $\zeta$ represents the axial coordinate 
$z \in \mathbb{R}$
when we consider a cylindrical geometry or the polar
angle $\theta \in [0,\pi)$
in the case of spherical symmetry. In the first case the metric elements depend  on $r_{\pm}$; in the second one, the coefficient $g^{\pm}_{\varphi\varphi}$ depends also on $\theta$. As usual, $t\in\mathbb{R}$ and $\varphi \in [0,2\pi)$. For wormholes, the normal coordinate associated to the radial direction is defined as
$d\eta = \pm \sqrt{g^{\pm}_{rr}} \, dr_{\pm}$,
i.e., ($\pm$)$\eta$ measures the perpendicular proper distance at the vicinities of the throat located at $\eta = 0$, for a static observer; the unit normal vector to the shell is correspondingly defined as $n_{\mu}  = \partial_{\mu} \eta \,$, pointing from $-$ to $+$. For an interior region joined to an exterior one, the shell is placed at  $\eta=\eta_0$ and $d\eta=+\sqrt{ g_{rr}^\pm }dr_\pm$.

Tides across shells are precisely described in terms of  the covariant relative acceleration usually expressed  as \cite{book,grav}
\be\label{A}
(\Delta a)^\mu =
- g^{\rho\mu}{R}_{\rho\alpha\nu\beta} V^\alpha (\Delta x)^\nu V^\beta,
\ee   
where ${R^\mu}_{\alpha\nu\beta}$ is the Riemann tensor, $V^\mu$ is the four-velocity of the travelling object, and $(\Delta x)^\mu$ is  the oriented  small separation of two points in spacetime. The Riemann tensor can be expressed as 
\be
R_{\mu\alpha\nu\beta} 
=  \Theta(-\eta) R^{-}_{\mu\alpha\nu\beta} + \Theta(\eta) R^{+}_{\mu\alpha\nu\beta} 
- \delta(\eta) 
\left[\kappa_{\alpha\beta} \, n_\mu n_\nu+\kappa_{\mu\nu} \, n_\alpha n_\beta-\kappa_{\alpha\nu} \, n_\mu n_\beta-\kappa_{\mu\beta} \, n_\alpha n_\nu\right] 
\ee
where $R^{\mp}_{\mu\alpha\nu\beta}$ is the smooth part of the tensor at each side of the shell and
\be
\kappa_{\alpha\beta} = \frac{1}{2} \left(\frac{\partial g^+_{\alpha\beta}}{\partial \eta}\Big|_{b_+} - \frac{\partial g^-_{\alpha\beta}}{\partial\eta}\Big|_{b_-} \right)
\ee
are the components of the jump in the extrinsic curvature tensor at the shell. 

With these definitions, we can write  the radial tide
on a radially moving object across the shell as in \cite{nos21},
\be\label{A2}
\Delta a_r=-
{\kappa^{0}}_{0}+\left[  {{R^{r0}}_{r0}}^{-}\Big|_{b_-} +{{R^{r0}}_{r0}}^{+}\Big|_{b_+} \right]
\frac{\Delta \tilde\eta}{2} 
+ \mathcal{O}(\Delta \tilde\eta^2) 
,
\ee
where ${\kappa^{0}}_0$ is the jump of the component ${K^{0}}_0$ of the extrinsic curvature across the infinitely thin shell, and $\Delta\tilde\eta$ is the proper radial separation between points of the object going through the shell; this reduces to $\Delta\eta$ for a rest body. The smooth regions contribute with a term proportional to $\Delta\tilde\eta$, 
which describes a tension acting on the body; this vanishes in the limit $\Delta\tilde\eta\to 0$, and in general does not constitute an important problem, even for non trivial topologies (see, for instance, \cite{martin}). The jump in the extrinsic curvature, instead, adds a finite contribution which is fixed for a given geometry: it does not vanish for infinitely close points at different sides of the surface $\Sigma$. This particular nature of the radial relative acceleration implies the practical problem of great tides acting on a body extended across the shell.

In an analogous way, we can write down the result in \cite{nos21} for the angular tide
of a radially moving object, which is conveniently decomposed as a divergent part proportional to the jump of the extrinsic curvature,  plus a finite one associated to the components of the Riemann tensor for the smooth regions:
\begin{eqnarray}  \label{Ap}
\Delta a_\perp & = & \Delta x_\perp\, \frac{\gamma \beta \,
{\kappa^{\varphi}}_{\varphi}}{\delta\tau}\nonumber\\
&  & +\,
\frac{\Delta x_{\perp}}{2} \,
\left[
{{R^{\varphi 0}}_{\varphi 0}}^-
+\gamma^2 \beta^2 \,
 \left(
 {{R^{\varphi 0 }}_{\varphi 0}}^- -  \, {{R^{\varphi r}}_{ \varphi r}}^-
\right)
\right]_{b_-}\nonumber\\
& & +\,\frac{\Delta x_{\perp}}{2}\left[{{R^{\varphi 0}}_{\varphi 0}}^+ +\gamma^2 \beta^2 \,
 \left(
 {{R^{\varphi 0 }}_{\varphi 0}}^+ -  \, {{R^{\varphi r}}_{ \varphi r}}^+
\right)
\right]_{b_+}.
\end{eqnarray}
In the expressions above we have introduced  the infinitely-short travelling proper time $\delta\tau$ of a point of the object across the shell.
Expressions with the coordinate $z$ instead of $\varphi$ hold for the axial tide in the case of a problem with symmetry along the $z$ axis. The parameters $\beta$ and $\gamma = 1/\sqrt{1-\beta^2}$ are defined through the velocity $V^{\hat{\mu}} = dx^{\hat{\mu}}/d\tau=(\gamma, \gamma \beta, 0, 0)$ measured in an orthonormal frame $\{ \vec{e}_{\hat{\mu}}\}$ at rest at the vicinities of the surface $\Sigma$. 

Differing from the case of radial tides, for tides in transverse directions to the radial one both contributions are proportional to the transverse extension $\Delta x_{\perp}$ of the object. The divergent character of the result is apparent because of the proper time $\delta\tau$, which goes to zero for an infinitely thin shell, in the   denominator of the first term in (\ref{Ap}). This serious difficulty could be avoided only in the case of a negligible velocity, or if the geometry is chosen so that it presents no  curvature jump at the throat. In what follows we show that this feature can be much easier to fulfil when shells connect inner to outer regions of topologically trivial spacetimes across a surface as long as the flare-out condition is not satisfied there (so we exclude the so-called ```bag of gold'' or ``baby universe'' geometries \cite{wheeler}). We will also find that cylindrical shells present an advantage as, even when composed by normal matter, they allow to  avoid troublesome tides in the transverse directions, while this is not possible in the case of spherical shells.

\section{Spherical shells}

We first consider spherically symmetric shells connecting inner to outer regions (with the restriction noted above) or two exterior submanifolds, depending on the existence of one or two asymptotic regions in the spacetime studied. The shell surface energy-momentum   tensor ${S^i}_j=\mathrm{diag}(-\sigma,p,p)$ is related with the jump of the metric derivatives at $\Sigma$ by the Lanczos equations \cite{sen,lanc,darm,isr}
\be  \label{le2}
8\pi {S^i}_j  = \kappa {\delta^i}_j - {\kappa^i}_j 
\ee
where $\kappa$ is the trace of the jump of the extrinsic curvature tensor. Hence the energy and pressure are given by
\begin{eqnarray}
\sigma & = & -\frac{1}{4\pi}{\kappa^\varphi}_\varphi\\
p & = & \frac{1}{8\pi}\lp{\kappa^0}_0+{\kappa^\varphi}_\varphi\rp.
\end{eqnarray}
On the other hand, the  contributions ${\cal T}_\alpha$ to the radial and transverse tides which generate traversability problems satisfy, according to Eqs. (\ref{A2}) and (\ref{Ap}), 
\be
{\cal T}_r\sim {\kappa^0}_0,\ \ \ \ \ \ \ \ \
{\cal T}_\perp\sim {\kappa^\varphi}_\varphi.
\ee
Hence we find relations between the energy-momentum of the shell and the troublesome 
radial ${\cal T}_r$ and transverse ${\cal T}_\perp$  terms which show that: 1) We could avoid the problem of the radial tide (which in general does not vanish for infinitely close radially oriented points) by a parameter choice such that ${\kappa^0}_0=0$; this does not cancel the transverse pressure, but only reduces it to a part proportional to ${\kappa^\varphi}_\varphi$. 2) The  divergent part of any transverse (angular) tide could be cancelled, but only at the price of admitting a shell with vanishing surface energy density.

We emphasize that these general features are not peculiar of any global behaviour of the geometry; their actual realization, however, requires a detailed analysis. We start from  general geometries of the form
\be
ds^2=- f(r)dt^2+ g(r)dr^2+ h(r)(d\theta^2+\sin^2\theta d\varphi^2).\label{metric}
\ee
We consider two submanifolds ${\cal M}_\pm$ described by metrics of this kind and connected by a spherically symmetric shell placed at the surface $\Sigma$ at $r_\pm=b_\pm$, with the signs $-$ and $+$ depending on the side of the shell; the relation between $b_-$ and $b_+$ comes from the condition of continuity of the metric across $\Sigma$, which is given by $h_-(b_-)=h_+(b_+)$. To include both a shell joining an inner to an outer submanifolds, as well as two exterior regions joined at a throat, we introduce the coefficient $\delta=\pm 1$, with $+1$ for trivial topologies, and $-1$ for a topologically non trivial geometry. Then
the jump of the components of the extrinsic curvature across the surface $\Sigma$ read
\be\label{kappat}
{\kappa^0}_0 =  \frac{f_+'(b_+)}{2f_+(b_+)\sqrt{g_+(b_+)}}-\delta\frac{f_-'(b_-)}{2f_-(b_-)\sqrt{g_-(b_-)}}
\ee
and
\be\label{kappafi}
{\kappa^\theta}_\theta =   {\kappa^\varphi}_\varphi =\frac{h_+'(b_+)}{2h_+(b_+)\sqrt{g_+(b_+)}}-\delta\frac{h_-'(b_-)}{2h_-(b_-)\sqrt{g_-(b_-)}}. 
\ee
The case of wormholes symmetric across the throat studied in Refs. \cite{nos21,nos22}  is recovered with $\delta =-1$ and identical functions $f$, $g$ and $h$ at both sides.  Note that the situation is rather different in each case:  while for little or null jump of the extrinsic curvature at a shell connecting two outer submanifolds very restrictive conditions should be imposed on the derivatives $f'$ and $h'$ at $b_\pm$, the same can be more easily achieved on a shell connecting an interior to an exterior region. In particular, for geometries with the reasonably desired behaviour of increasing areas associated to increasing radial coordinates, that is $h_\pm'>0$, for a shell joining two outer submanifolds we would have the sum of two positive terms in Eq. (\ref{kappafi}) above; hence the condition ${\kappa^\varphi}_\varphi=0$ allowing to solve the difficulties with transverse tides (apart from the resulting exotic nature of the shell matter\footnote{Recall that a null energy density implies that just one negative pressure (positive tension) would imply the violation of the null, weak and strong energy conditions.}) can only be demanded in the case of an inner region joined to an outer --infinite-- one.  

\subsection{Example: shells connecting string cloud geometries}

As an example with more than a single parameter (as it would be in the case of the Schwarzschild black hole), and thus with the flexibility  to achieve the general conditions established, we consider a mass $M$ in the center of a spherically symmetric cloud of strings. The cloud  energy-momentum tensor reads
\be
T^{\mu\nu}=\frac{\rho_0}{\sqrt{-g_S}}\epsilon^{ln}\frac{\partial x^\mu}{\partial \lambda^l}\frac{\partial x^\nu}{\partial \lambda^n}
\ee
with $\rho_0$ the proper density of the cloud,  $\epsilon^{ln}$ the two dimensional Levi--Civita symbol and $g_S$ the determinant of the induced metric $g_{ln}$ on the string world sheet parametrized by $\lambda^0$ and $\lambda^1$.
The cloud gauge invariant density is $\rho_0\sqrt{(-g_S)}=A/r^2$, with $A$ a positive constant; see Ref.  \cite{cloud} and also \cite{barvil,dangre,ram,martin}. If two submanifolds  associated to such kind of matter with different parameters $M$ and $A$ are joined at a surface $\Sigma$, the metric at each side of it has the form 
\be
ds_\pm^2=-f_\pm(r_\pm)dt_\pm^2+f_\pm^{-1}(r_\pm)dr_\pm^2+r_\pm^2\left(d\theta^2+\sin \theta d\varphi^2\right)
\ee
with 
\be
f_\pm(r_\pm)=1-A_\pm-\frac{2M_\pm}{r_\pm}.
\ee
For parameters $0<A_\pm<1$ the geometry at each side would have one horizon located at
\be
R_\pm=\frac{2M_\pm}{1-A_\pm}.
\ee
We will not deal with the case $A_\pm>1$ (homogeneous spacetimes) and we assume $M_\pm>0$ (so the possibility of  no horizons and hence a naked singularity at $r=0$ is ruled out). Given the form of the metric, the continuity condition $h_-(b_-)=h_+(b_+)$ implies $b_-=b_+$. Then we consider a shell at a surface $\Sigma$ of radius $b$ joining the submanifolds ${\cal M}_-$ and ${\cal M}_+$, with the restriction $b> R_{\rm Max}$, where ${R_{\rm  Max}}$ is the largest of the two horizon radii $R_\pm$. In the case $\delta=-1$ (thin-shell wormhole studied in \cite{martin}) this condition allows a traveller to traverse across the wormhole finding no  horizons in each side;   in the case $\delta=+1$ --a shell joining an interior region with an exterior one-- this implies that a traveller can reach the shell coming from outside  without finding an horizon before experiencing the troublesome tides across the surface $\Sigma$. 

For a wormhole (in general, not symmetric across the throat) the jump of the component ${K^0}_0$ of the extrinsic curvature reads
\be
{\kappa^0}_0= \frac{M_+}{b^2\sqrt{1-A_+-2M_+/b}}+\frac{M_-}{b^2\sqrt{1-A_--2M_-/b}}.
\ee 
Thus in the case of a non trivial topology it is clearly impossible to fulfil the condition of a null jump of the extrinsic curvature; no solution is then possible for the difficulties with the radial tide.  
For a shell joining an interior region to an exterior one, instead, we have
the jump of the extrinsic curvature
\be
{\kappa^0}_0= \frac{M_+}{b^2\sqrt{1-A_+-2M_+/b}}-\frac{M_-}{b^2\sqrt{1-A_--2M_-/b}}.
\ee 
Now it is clear that the possibility of ${\kappa^0}_0=0$ is not excluded, except in the Schwarzschild limit of both $A_-$ and $A_+$ equal to zero.
After some elementary algebra and recalling the definition of the horizon radii $R_-$ and $R_+$, for a given set of parameters characterizing the geometries at each side of the shell the solution for the condition of vanishing curvature jump is given by
\be
b^*=\frac{\lp M_+-M_-\rp R_+R_-}{M_+R_+-M_-R_-}.
\ee
This solution makes sense if $b^*$ is larger than $R_{\rm Max}$. Assume that  $R_+>R_-$ (so that $R_{\rm Max}=R_+$, which is consistent with being the submanifold ${\cal M}_+$ the outer one),  $M_+<M_-$\footnote{The possibility of a larger horizon radius associated to a smaller mass is given by the additional parameters $A_\pm$ describing the string clouds densities at each side.} and $M_-R_->M_+R_+$, which ensures $b^*>0$. Then we can write $M_+=\alpha_m M_-$, $R_+=\alpha_r R_-$, with $0<\alpha_m<1$, $\alpha_r>1$ and $\alpha_m\alpha_r<1$; therefore the expression for the shell radius $b^*$ turns into
\be
b^*=\frac{\alpha_r(\alpha_m-1)}{\alpha_r\alpha_m-1}R_-=\frac{(\alpha_m-1)}{\alpha_r\alpha_m-1}R_+.
\ee
We immediately see that the conditions on the coefficients $\alpha_m$ and $\alpha_r$ imply $\alpha_r(\alpha_m-1)/(\alpha_r\alpha_m-1)>0$ and $(\alpha_m-1)/(\alpha_r\alpha_m-1)>1$; hence the requirement $b^*>R_{\rm Max}$ of a shell placed beyond the largest horizon radius  is satisfied. In other words, in a spacetime with only one asymptotic region, given a set of parameters $M_-$, $M_+$, $A_-$ and $ A_+$ defining geometries with horizon radii $R_-$ and $R_+$, a shell radius can be selected in order to cancel the jump of the component ${K^0}_0$ of the extrinsic curvature, and then the problem with the radial tide on a radially moving object can be avoided.

On the other hand, the  transverse components 
\be
{\kappa^\varphi}_\varphi
= {\kappa^\theta}_\theta =
\frac{\sqrt{1 - A_+ - 2 M_+ / b}}{b} -\frac {\sqrt{1 - A_- - 2 M_-/ b}}{b}
\ee
can be cancelled out with a shell radius which, with the same definitions above, is given by 
\be
b^*=\frac{(M_+ - M_-) R_- R_+}{M_+ R_- - M_- R_+}.
\ee
 Again, this solution makes sense if $b^*$ is larger than $R_{\rm Max}$. If we assume $R_{\rm Max}=R_+>R_-$,  $M_+>M_-$ and $M_+R_- > M_-R_+$, which ensures $b^*>0$, then we can write $M_- = \alpha_m M_+$ and $R_+=\alpha_r R_-$ with $0<\alpha_m<1$, $\alpha_r>1$ and $\alpha_m\alpha_r<1$; thus the expression for the shell radius $b^*$ can be put in the form
\be
b^*=\frac{\alpha_r (\alpha_m-1) R_-}{\alpha_m - \alpha_r}
=\frac{(\alpha_m-1) R_+}{\alpha_m - \alpha_r}
\ee
and hence the requirement $b^*> R_{\rm Max}$ is satisfied. With this choice the component of the curvature jump associated to the radial tide would be negative: ${\kappa^0}_0|_{b^*} < 0$. Consequently, the shell would have null surface energy density and negative transverse pressures (positive tensions).

\section{Cylindrical shells}

Now we consider cylindrically symmetric shells in both topologically trivial and non trivial spacetimes. The Lanczos equations relating the shell energy density $\sigma$ and pressures $p_\varphi$ and $p_z$ with the jump of the extrinsic curvature at the surface $\Sigma$ dictate
\begin{eqnarray}
\sigma & = & -\frac{1}{8\pi}\lp{\kappa^\varphi}_\varphi+{\kappa^z}_z\rp,\\
p_\varphi & = & \frac{1}{8\pi}\lp{\kappa^0}_0+{\kappa^z}_z\rp,\\
p_z & = &\frac{1}{8\pi}\lp{\kappa^0}_0+{\kappa^\varphi}_\varphi\rp.
\end{eqnarray}
The contributions to the radial and transverse tides which imply difficulties with traversability satisfy the following relations with the components of the extrinsic curvature jump:
\be
{\cal T}_r\sim {\kappa^0}_0,\ \ \ \ \ \ \ \ \
{\cal T}_{\perp,\varphi}\sim {\kappa^\varphi}_\varphi,\ \ \ \ \ \ \ \ \ {\cal T}_{\perp,z}\sim {\kappa^z}_z.
\ee
Thus we find that now the situation is, in a sense, better than in the spherical problem; the energy-momentum of the matter thin layer and the troublesome terms of the tides are related in such a way that:
1) Any of the two transverse formally divergent tides can be avoided, though this cannot be achieved simultaneously for both  ones, unless we accept a shell with a null energy density.
2) It would be possible, in principle, to simultaneously eliminate the difficulties with  radial and  angular tides, or with radial and longitudinal tides. This could be achieved by a choice compatible with a non vanishing energy density, which would imply a sort of ``directional dust'', i.e. matter such that one of the pressures would be zero. 

As before, note that no topology assumption has been made up to this point; however, actual difficulties can be very different in each case. In what follows we examine the possible realization of such conditions by studying  radial and transverse (angular and axial) tides across shells connecting  geometries of the form
\begin{equation}
ds^2 = -f(r)dt^2 +g(r)dr^2 +h(r)d\varphi ^2+k(r)dz^2,
\label{e1}
\end{equation}
where $f$, $g$, $h$ and $k$ are positive functions.  Let us write the jump of the components of the extrinsic 
curvature across a shell with radial coordinate $b_-$ or $b_+$ (depending of the side considered). The continuity of the metric implies the condition $h_-(b_-) = h_+(b_+)$, while in the longitudinal direction we simply must require $\sqrt{k_-(b_-) } dz_- = \sqrt{k_+(b_+) } dz_+$. Introducing the coefficient $\delta=\pm 1$ with the same physical meaning of the preceding section, we have \begin{eqnarray}
{\kappa^0}_0 & = & \frac{f_+'(b_+)
}{2f_+(b_+) \sqrt{g_+(b_+)}}-\delta\frac{f_-'(b_-)
}{2f_-(b_-) \sqrt{g_-(b_-)}},\\
{\kappa^\varphi}_\varphi & = &  \frac{h_+'(b_+)}{2h_+
(b_+)\sqrt{g_+(b_+)}}-\delta\frac{h_-'(b_-)}{2h_-
(b_-)\sqrt{g_-(b_-)}},\label{kappafi2}\\
{\kappa^z}_z & = & \frac{k_+'(b_+)}
{2k_+(b_+) \sqrt{g_+(b_+)}}-\delta\frac{k_-'(b_-)}
{2k_-(b_-) \sqrt{g_-(b_-)}}.
\end{eqnarray}
The results in \cite{nos21} for shells at cylindrical surfaces  satisfying a flare-out condition (cylindrical thin-shell wormholes) are recovered for $\delta=-1$ and the choice of equal functions $f$, $g$, $h$ and $k$ at each side. In the same fashion as in the spherical problem, we note the additional freedom available to choose the parameters reducing or eliminating the problematic contributions to tides for $\delta =+1$, that is, for shells connecting interior to exterior regions (under the same restriction above, excluding --now cylindrical-- ``bag of gold'' geometries). In particular, when $\delta =+1$ we would be able to achieve favourable conditions which are strictly impossible when $\delta=-1$: analogously as noted in the spherical case, consider two geometries satisfying the reasonable conditions $h_\pm'>0$, which ensure that cylinders increase their circumferences as the radial coordinates at each side grow. If we match the inner part of one of them with the outer part of the other, the problem with the angular transverse tide at the shell could be solved imposing the condition ${\kappa^\varphi}_\varphi=0$; but if two outer regions of such geometries are matched on a shell, such improvement for transverse tides is not possible because the two terms in the right hand side of Eq. (\ref{kappafi2}) would be positive. 

\subsection{Example 1: Shells connecting Levi--Civita geometries}

Consider a cylindrical shell at the spacelike surface where two submanifols with metrics of the Levi--Civita form
\be
ds^2 =r_\pm^{2d_\pm(d_\pm-1)}(-dt_\pm^2+dr_\pm^2)+W_\pm^2 r_\pm^{2(1-d_\pm)}d\varphi^2+r_\pm^{2d_\pm}dz_\pm^2
\ee
are connected\footnote{Dimensions are consistent as long as in this case the radial coordinate $r$ is understood as $\rho/\rho^*$, where $\rho^*$ is a ``core'' radius determining the scale. We consider $\rho>\rho^*$, so that $r>1$; thus we work in terms of a dimensionless line element $ds^2=(d\tilde{s}/\rho^*)^2$. The physical meaning of the parameters $W_\pm$, which can be assumed to be positive definite, is best understood in the conical case, where $2\pi(1-W_\pm)$ are the angle deficits in one turn around the symmetry axis.}. The parameters $d_\pm$ can be related to the mass per unit length; the case $d_\pm=0$  describes conical (gauge  cosmic string) spacetimes \cite{vilrep,vilbook}. In order to start the mathematical construction from geometries with reasonable asymptotic behaviours, we will restrict to positive $d_\pm$ to avoid a vanishing longitudinal length at radial infinity, and we assume $d_\pm<1$ in order to avoid  a circumference decrease for increasing $r_\pm$. While in this example, in the wormhole case, the areal definition of the flare-out condition is automatically fulfilled, such assumption ensures that also the circular definition of the flare-out condition is satisfied at the throat (see \cite{brle} and also the related works \cite{eisi10,bbs,bko}).

The components of the extrinsic curvature jump associated to the existence of an infinitely thin shell placed at $r_\pm=b_\pm$ read
\begin{eqnarray}
{\kappa^0}_0 & = & \frac{d_+(d_+-1)}{b_+^{d_+(d_+-1)+1}}-\delta \frac{d_-(d_--1)}{b_-^{d_-(d_--1)+1}}\,,\label{00}\\
{\kappa^\varphi}_\varphi & = & \frac{1-d_+}{b_+^{d_+(d_+-1)+1}}-\delta \frac{1-d_-}{b_-^{d_-(d_--1)+1}}\,,\label{fifi}\\
{\kappa^z}_z & = & \frac{d_+}{b_+^{d_+(d_+-1)+1}}-\delta \frac{d_-}{b_-^{d_-(d_--1)+1}}\,\label{zz}
\end{eqnarray} 
with $\delta=1$ for trivial topologies and $\delta=-1$ for non trivial ones.
The geometry continuity at the shell dictates the relation
\be\label{cont}
W_+b_+^{1-d_+}=W_-b_-^{1-d_-}
\ee
if we recall that we take $W_\pm>0$.  From these expressions we deduce that:  1) For conical geometries at both sides $(d_+=d_-=0)$, the conditions ${\cal T}_r=0$ and ${\cal T}_{\perp,z}=0$ can be simultaneously satisfied for both possible kinds of global topology (this agrees with the qualitative insight advanced in Ref. \cite{nos21}). Instead, the condition ${\cal T}_{\perp,\varphi}=0$ cannot be fulfilled for wormholes, as we would have two positive terms; on the other hand, for a geometry with only one asymptotic region it would imply $b_+=b_-$, which together with the continuity relation (\ref{cont}) would lead to $W_+=W_-$, thus eliminating the shell itself. 2)  If we discard conical metrics but, for the reasons established above, we then restrict to Levi--Civita spacetimes with $0<d_\pm<1$, in  the case of wormholes $(\delta=-1)$ equal signs are forced for both terms of  equations (\ref{00}), (\ref{fifi}) and (\ref{zz}); hence no component of the curvature jump can be cancelled, and no troublesome component of tides can be avoided. 3) Again for non conical submanifolds, for inner regions joined to outer ones $(\delta=1)$ each condition leading to cancel the curvature jump corresponding to any problematic tide could be satisfied; this means a clear advantage of trivial topologies over non trivial ones.  However, if we try to simultaneously cancel any two components of the extrinsic curvature jump, we are led to $d_+=d_-$, and in turn we would be forced to have $b_+=b_-$, which together with the continuity requirement (\ref{cont}) gives $W_+=W_-$; this means no different geometries, and no shell at all. In other words, for a shell connecting inner and outer Levi--Civita (non conical) submanifolds, only one of the three principal orientations could be fixed to be safe at once.

\subsection{Example 2: Shells connecting black string geometries}

Consider two submanifols with metrics of the form 
\be\label{bs1}
ds^2 =
- f_\pm(r_\pm) \, dt_\pm^2 
+ \frac{1}{f_\pm(r_\pm)} \, dr_\pm^2 
+ r_\pm^{2} d\varphi^2 
+ \alpha^2  r_\pm^{2} dz_\pm^2
\ee
with \cite{lemos}
\be\label{bs2}
f_\pm(r_\pm) = \alpha^2 r_\pm^{2} - \frac{4m_\pm}{\alpha r_\pm} + \frac{4 \lambda_\pm^2}{\alpha^2 r_\pm^2}.
\ee
Here $\alpha^2= -\Lambda /3 > 0$ is related to the cosmological constant, which we assume to be negative and the same for both submanifolds, and $m_\pm$ and  $\lambda_\pm$ are respectively the masses and charges per unit length. The charges are associated to electric fields of the form
\be
A_\mu=\lp-\frac{2\lambda}{\alpha r}+\mathrm{const},0,0,0\rp.
\ee
Such geometries are singular at the axis of symmetry (where the Kretschmann scalar diverges). Depending on the values of the parameters, metrics defined by (\ref{bs1}) and (\ref{bs2}) can have $f=0$ and present an event horizon; thus the {\it black string} denomination. When $m=0$ and $\lambda =0$ no horizons are possible, while if $m \neq 0$ and $\lambda =0$, there is only an event horizon located at $\alpha r_h = (4m)^{1/3}$. In the case $\lambda \neq 0$ the horizons are given by the positive roots of the fourth degree polynomial $P(r) = \alpha ^4 r^4 - 4m \alpha r + 4 \lambda^2$; if $0<|\lambda | < \lambda_e = m^{2/3}\sqrt{3}/2$ there are two horizons satisfying $\alpha r_{-} < \alpha r_{+} < (4m)^{1/3}$, with $r_{+}$ the radius of the event one, while for the extremal charge per unit length $|\lambda | = \lambda _e$  only one horizon with radius $\alpha r_e = m^{1/3}$ exists; finally if $|\lambda | > \lambda _e $ there are no horizons, and we have only a naked singularity at the axis of symmetry.

If we join submanifolds described by the metrics above (see Ref. \cite{eec}), the geometry continuity at the shell dictates the relation $r_\pm = b_\pm = b$. The components of the extrinsic curvature jump associated to the existence of an infinitely thin shell placed at $b$ read
\begin{eqnarray}
{\kappa^0}_0 & = & 
\frac{2 \, b \, m_+ \alpha + b^4 \alpha^4 - 4 \, \lambda_+^2}{b^2 \, \alpha \sqrt{ b^4 \alpha^4 - 4 \,b \, m_+ \, \alpha  + 4 \, \lambda_+^2}}
-\delta 
\frac{2 \, b \, m_- \alpha + b^4 \alpha^4 - 4 \, \lambda_-^2}{b^2 \, \alpha \sqrt{ b^4 \alpha^4 - 4 \,b \, m_- \, \alpha  + 4 \, \lambda_-^2}}
\,,\label{00bs}\\
{\kappa^\varphi}_\varphi & = &
{\kappa^z}_z = 
\frac{\sqrt{ b^4 \alpha^4 - 4 \,b \, m_+ \, \alpha  + 4 \, \lambda_+^2}}{b^2 \, \alpha }
-\delta 
\frac{\sqrt{ b^4 \alpha^4 - 4 \,b \, m_- \, \alpha  + 4 \, \lambda_-^2}}{b^2 \, \alpha }
\,,\label{fifibs}
\end{eqnarray} 
with $\delta=1$ for trivial topologies and $\delta=-1$ for non trivial ones. In the case of trivial topologies, the transverse components vanish in general for
\be
b^*= \frac{\lambda_+^2 - \lambda_-^2}{\alpha(m_+ - m_-)} ,
\ee
while 
\be
{\kappa^0}_0 |_{b^*}
= 
2\, \alpha \,
\frac{(m_+-m_-)^4}{\lambda_-^2 - \lambda_+^2} \,
\left[ 4 \, (m_+-m_-)^3 (m_+ \, \lambda_-^2 - m_- \, \lambda_+^2 ) + (\lambda_+^2 - \lambda_-^2 )^4 \right]^{-1/2}
\ee
is non-null for finite $b^*$. We must demand $b^*$ to be greater than the event horizons of the interior and exterior geometries, or start from geometries without event horizons and singularities in order to have an accessible shell and avoid any problem. We can check the special case of a shell around a vacuum bubble, i.e. with $m_-=0$ and $\lambda_- = 0$, so that the interior metric has no horizons nor singularities and $b^*_{\rm bubble} = \lambda_+^2 / (\alpha\, m_+)$. In particular, to avoid event horizons in this case we must work with $|\lambda_+| > \lambda_e = m_+^{2/3} \sqrt{3}/2$, where $\lambda_e $ is the extremal charge of the exterior geometry. These vacuum bubbles have ${\kappa^\varphi}_\varphi = 0 = {\kappa^z}_z$ and ${\kappa^0}_0 = - 2 \alpha \, m_+^4 / \lambda_+^6$, which allows to avoid troublesome transverse tides across the shell, though it is in correspondence  with a null energy density and negative pressures (positive tensions) on the surface of the shell.

\section{Summary}

We have investigated the traversability across thin shells in relation with the difference in tides associated to different global properties of the geometries. We have worked under the assumption of topologically trivial spacetimes without a throat (a surface satisfying a flare-out condition), or non topologically trivial spacetimes  with a throat  connecting two asymptotic regions; we have devoted special attention to the character of matter on the shells at the surfaces joining different submanifolds, i.e. the possibility --or not-- of avoiding null energy densities and, in general, exotic matter. Our analysis has been carried out in a generic form valid for spherical and cylindrical static configurations, but we have provided examples illustrating different aspects and related difficulties. We have shown that, while curvature jumps across infinitely thin shells lead to troublesome tides across a throat in spacetimes with two asymptotic regions, associated contributions of tides across shells joining inner with outer submanifolds of spacetimes with one asymptotic region can be significantly reduced or even  cancelled. Besides we have found an important difference between spherically symmetric shells and cylindrically symmetric ones: for matter with positive energy density, the first ones are not compatible with vanishing  troublesome contributions to transverse tides, and only admit a solution for the radial problem; the second ones, instead, allow to avoid the (studied here) traversability problems for objects extended along any of the three principal directions, though not simultaneously for both transverse directions.

\end{document}